\documentclass[twocolumn,prl,aps,superscriptaddress,showpacs,preprintnumbers]{revtex4}

\usepackage{graphicx}

\begin{document}

\title{ Two characteristic energies in the low-energy antiferromagnetic response of the
electron-doped high-temperature superconductor Nd$_{2-x}$Ce$_x$CuO$%
_{4+\delta}$ }
\author{G.~Yu}
\affiliation{Department of Physics, Stanford University, Stanford, California 94305}
\author{Y.~Li}
\affiliation{Department of Physics, Stanford University, Stanford, California 94305}
\author{E.M.~Motoyama}
\affiliation{Department of Physics, Stanford University, Stanford, California 94305}
\author{K.~Hradil}
\affiliation{Institut f\"ur Physikalische Chemie, Universit\"at G\"ottingen, 37077
G\"ottingen, Germany }
\author{R.A.~Mole}
\affiliation{Forschungsneutronenquelle Heinz Maier-Leisbnitz, 85747 Garching, Germany}
\author{M.~Greven}
\affiliation{Department of Applied Physics, Stanford University, Stanford, California
94305}
\date{\today}

\begin{abstract}
Inelastic neutron scattering for Nd$_{2-x}$Ce$_x$CuO$_{4+\delta}$ near
optimal doping ($x \approx 0.155$, $T_{c} = 25 \, \mathrm{K}$) reveals that
the dynamic magnetic susceptibility at the antiferromagnetic zone center exhibits
two characteristic energies in the superconducting state: $\omega_1
\approx 6.4\, \mathrm{meV}$ and $\omega_2 \approx 4.5 \, \mathrm{meV}$. 
These two magnetic energies agree $quantitatively$ with the 
$B_{1g}$/$B_{2g}$ and $A_{1g}$ features previously observed in electronic Raman
scattering, where the former is believed to indicate the maximum electronic gap 
and the origin of the smaller $A_{1g}$ feature has remained unexplained. The
susceptibility change upon cooling into the superconducting state is
inconsistent with previous claims of the existence of a magnetic resonance
mode near 10 meV, but consistent with a resonance at $\omega_2$.
\end{abstract}

\pacs{74.72.Jt,74.25.Ha,78.70.Nx}
\maketitle

Antiferromagnetic (AF) fluctuations might contribute to  
the superconducting (SC) pairing in the cuprate
high-temperature superconductors. 
The most prominent magnetic feature observed in the SC state is the 
$resonance$, an unusual spin-triplet ($S=1$) collective mode centered at the
two-dimensional AF zone center $\mathbf{Q}_{AF}=(0.5,0.5)$ r.l.u. 
\cite{EschrigAdvPhys06}. The resonance has been observed in inelastic neutron
scattering (INS) experiments on several families of hole-doped cuprates: YBa$%
_{2}$Cu$_{3}$O$_{6+\delta }$ \cite{Rossat-MignodPhysica91} and Bi$_{2}$Sr$%
_{2}$CaCu$_{2}$O$_{8+\delta }$ \cite{FongNature99}, which are 
comprised of two CuO$_{2}$ layers per unit cell, as well as in
single-layer Tl$_{2}$Sr$_{2}$CuO$_{6+\delta }$ \cite{HeScience02} and
HgBa$_{2}$Cu$_{2}$O$_{4+\delta }$ \cite{Guichuan08}. The origin of the
resonance and of its unusual dispersion has been a topic of much recent
debate \cite{EschrigAdvPhys06, Carlson2004, Kruger2004, PrelovsekSega2006,
Seibold2006, Uhrig2004, Vojta2006}. 

Magnetic INS measurements of the electron-doped compounds have become
possible only in recent years. For Nd$_{2-x}$Ce$_{x}$CuO$_{4+\delta }$
(NCCO), such measurements have revealed a gap below $T_{c}$ \cite{YamadaPRL03},
the effect of a magnetic field on this gap \cite{MotoyamaPRL06}, and the
evolution with doping and temperature of the spin-correlations in the Cu-O
layers \cite{MangPRL04, MotoyamaNature07}. These measurements were possible
in crystals of sufficiently large size and quality, despite the fact that
the SC compounds exhibit chemical inhomogeneities and a secondary chemical
phase \cite{MangPRB04}. 
Results for electron-doped Pr$_{0.88}$LaCe$_{0.12}$CuO$%
_{4}$ ($T_{\mathrm{c}}=24\,\mathrm{K}$) \cite{WilsonNature06} and NCCO ($T_{%
\mathrm{c}}=25\,\mathrm{K}$) \cite{ZhaoPRL07} were interpreted as indicative
of a magnetic resonance with an energy of about 10 meV, suggesting that the
resonance may be a universal feature of the cuprates, independent of the
type of carriers.

In contrast to INS, which provides information about the magnetic degrees of freedom, 
electronic Raman scattering yields information about the
charge dynamics. Polarization analysis has led to the identification of
several characteristic energies in both hole- and electron-doped cuprates 
\cite{Devereaux2007,QazilbashPRB05}. Features observed in $B_{1g}$ and 
$B_{2g}$ symmetry have been associated with the normal state pseudogap
or the SC gap, whereas the unexpected
observation of a feature in $A_{1g}$ symmetry has found no
widely-accepted explanation. One suggestion for the hole-doped compounds 
is that the latter may
be associated with the magnetic resonance \cite{Gallais2002, Letacon2005, Letacon2006JPCS}. 

In this Letter, we report a detailed INS study of the low-energy dynamic
magnetic susceptibility $\chi ^{\prime \prime }(\mathbf{Q},\omega )$ of NCCO
near optimal doping (onset $T_{\mathrm{c}}=25\,\mathrm{K}$). 
As the system is cooled into the SC state,
$\chi ^{\prime \prime }(\mathbf{Q}_{AF},\omega )$ 
exhibits a spectral weight shift from below to above $\omega _{1}=6.4(3)$ meV, 
which can be described as the opening of a gap, as
well as a local maximum at $\omega _{2}=4.5(2)$ meV, below the gap.
Remarkably, these two energies agree $quantitatively$ with those
obtained from Raman scattering in $B_{1g}$/$B_{2g}$ and $A_{1g}$ symmetry,
respectively \cite{QazilbashPRB05}. The larger of the two corresponds to the
maximum $2\Delta _{el}$ of the non-monotonic electronic $d$-wave gap \cite%
{QazilbashPRB05,MatsuiPRL05}, whereas the lower energy scale likely
indicates the presence of a resonance, consistent
with the situation for the hole-doped cuprates for which the resonance is always found 
below $2\Delta _{el}$. This is supported
by our measurement of the temperature dependence of the susceptibility 
at $\omega _{2}$,
which reveals a monotonic increase upon cooling
into the SC state. The present results, while overall consistent with prior
data for electron-doped compounds, do not support the claim of the existence
of a magnetic resonance at higher energies \cite{WilsonNature06,ZhaoPRL07}.

Two NCCO crystals were grown in a traveling-solvent floating-zone furnace in
an oxygen atmosphere of $5\,\mathrm{bar}$. To remove the excess oxygen and
achieve superconductivity, the crystals were annealed for 10 h in flowing
argon at $970^{\circ }\mathrm{C}$, followed by 20 h in flowing oxygen at $%
500^{\circ }\mathrm{C}$. The Ce concentration was carefully measured by
inductively coupled plasma (ICP) atomic emission spectrometry on several
parts cut from the crystals. 
The Ce concentration was found to vary both along the diameter of the sample
and along the growth direction.
For the primary crystal
used in this study (diameter: 4 mm; mass: 6.2 g), we estimate the overall
composition to be $x=0.157(7)$. As discussed below, the chemical
inhomogeneity manifests itself as a broadening of the features observed in
our experiment. The value of $T_{c}$ was determined from magnetic susceptibility
measurements of two small pieces (with compositions $x\approx 0.150$ and
0.164) cut from the ends of the crystal. Despite the somewhat different
composition of the end pieces, the onset temperature of the transition is
nearly the same, consistent with the maximum value of $T_{\mathrm{c}}=25\,%
\mathrm{K}$ generally obtained at and near optimal doping. The second
crystal was smaller (mass: 5 g) and has a nearly identical composition [$%
x=0.156(4)]$ and the same value of $T_{c}$. 

The INS experiment was performed on the thermal triple-axis instrument PUMA
at the FRM-II in Garching, Germany. The two samples were mounted in 
separate measurements inside a low-temperature displex such that the 
$(H,K,0)$ plane was parallel to the horizontal scattering plane. We used a double-focusing PG(002)
monochromator, a focusing PG(002) analyzer, and a fixed final energy of $%
14.7\,\mathrm{meV}$ with a PG filter after the sample. The energy resolution
ranged from about 0.8 to 1.4 meV (FWHM) between $\omega =1.5$ and 12 meV.
The room temperature lattice constants are $a=3.93\,\mathrm{%
\mathring{A}}$, $c=12.08\,\mathrm{\mathring{
A}}$. Using a horizontally
flat monochromator and analyzer, a rocking scan at the $(2,0,0)$ reflection
indicated a mosaic of $0.3^{\circ }$ (FWHM).

\begin{figure}[t]
\includegraphics[width=3.5in]{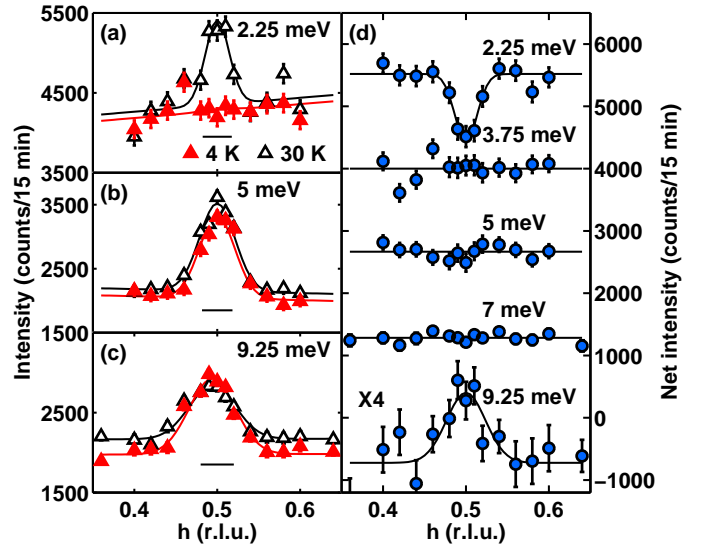}
\caption{(a-c) Representative $[h, 1-h, 0]$ momentum scans through the AF
zone center at $T=4$ and 30 K. The lines represent fits to a Gaussian. The
horizontal bars indicate the resolution. (d) Intensity difference between $T
= 30$ and 4 K. For clarity, data sets are shifted relative to each other by
1400 counts/15 min and the difference at $\protect\omega = 9.25$ meV is
multiplied by a factor of four. }
\label{fig2}
\end{figure}

Figure \ref{fig2}(a)-(c) shows representative $[h, 1-h, 0]$ transverse
momentum scans below ($4\, \mathrm{K}$) and above ($30\, \mathrm{K}$) $T_%
\mathrm{c} $. At all energy transfers, the magnetic response remained
centered at $\mathbf{Q}_{AF}$. Figure 1(a) demonstrates that at $%
\omega = 2.25$ meV the normal state response is resolution-limited and the
magnetic scattering disappears completely deep in the SC state, signaling
the opening of a gap. On the other hand, at $\omega = 9.25$ meV, in addition
to an overall change in background scattering, a clear enhancement at $%
\mathbf{Q}_{AF}$ is observed in the SC state [Fig.~1 (c) and (d)]. Between
3.75 and 7 meV, the intensity difference is featureless, as seen from Fig.~1(d).

\begin{figure}[t]
\label{fig3}\includegraphics[width=3.5in]{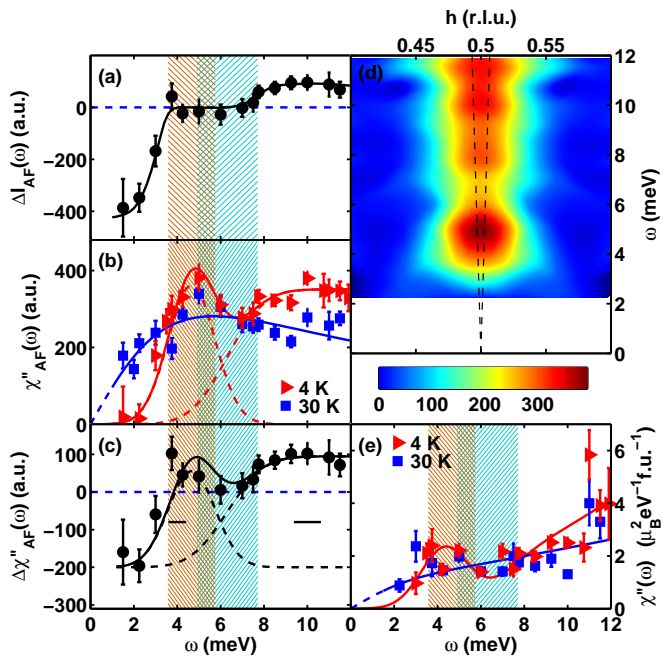}
\caption{(a) Difference between $T = 4$ and 30 K of the intensity amplitude
at the antiferromagnetic wave vector $\mathbf{Q}_{AF}$. (b) Peak
susceptibility $\protect\chi^{\prime\prime}_{AF}(\protect\omega)$, obtained
by correcting the measured peak intensity for the Bose factor. (c) Relative
change in $\protect\chi^{\prime\prime}_{AF}(\protect\omega)$ between $T = 4$
and 30 K. (d) Contour plot of $\protect\chi^{\prime \prime }(\mathbf{Q},%
\protect\omega )$ at 4 K, made by interpolation of symmetrized momentum
scans of the kind shown in Fig.~1 (a)-(c), with a constant background
removed. (e) Local susceptibility, obtained in absolute units from the
momentum-integral of $\protect\chi^{\prime \prime }(\mathbf{Q},\protect%
\omega )$ by comparing with the measured intensity of acoustic phonons. The
continuous lines in (a) and (e) are guides to the eye, while the lines in
(b) and (c) are the results of fits, as described in the text. The
horizontal bars in (c) indicate the FWHM energy resolution of 0.9 and 1.25
meV at $\protect\omega = 4$ and 10 meV, respectively. The shaded vertical
bands centered at about 4.5 and 6.5 meV indicate the respective ranges of $%
A_{1g}$ and $B_{1g}$ peak energies from Raman scattering \protect\cite%
{QazilbashPRB05} corresponding to the chemical inhomogeneity in our sample.
The INS data were obtained during two separate experimental runs. Peak
intensities and susceptibilities were obtained from Gaussian fits. The
energy dependence of the intrinsic momentum width is approximately linear
above 4 meV, with a slope of 320(30) meV\AA, and indistinguishable
between 4 K and 30 K. Assuming a cone of spin-wave-like excitations, we
estimate the corresponding velocity to be 175(50) meV\AA,
significantly smaller than the spin-wave velocity of about 1 eV\AA~
in undoped Nd$_2$CuO$_4$ [dashed line in (d)]
\protect\cite{BourgesPRL97}. }
\end{figure}

Figure 2(a) reveals a clear shift of the intensity from below 4 to above 7
meV upon cooling. The peak susceptibility 
$\chi _{AF}^{\prime \prime }(\omega ) \equiv \chi ^{\prime \prime }(\mathbf{Q}_{AF},\omega )$
at 4 and 30 K and the susceptibiliy difference between the two temperatures
are shown in Fig.~2(b) and (c), respectively. 
Due to the change in Bose factor, the local maximum in the
susceptibility difference directly results from the observed zero (within
error) net peak intensity between 3.75 and 7 meV. The normal state response
at 30 K is quite well described by a Lorentzian, $\chi^{\prime\prime}_{AF}(%
\omega)= \chi^{\prime\prime}_{AF}\Gamma\omega/(\Gamma^2 + \omega^2)$, with
relaxation rate $\Gamma = 5.7(5)$ meV. In contrast, the excitation spectrum
in the SC state exhibits two characteristic energies: a maximum at 4-5
meV and a minimum at 6-7 meV. This can also be seen from the susceptibility
difference in Fig.~2(c) and from the low-temperature contour plot in Fig.~2(d).

\begin{figure}[t]
\includegraphics[width=2.3in]{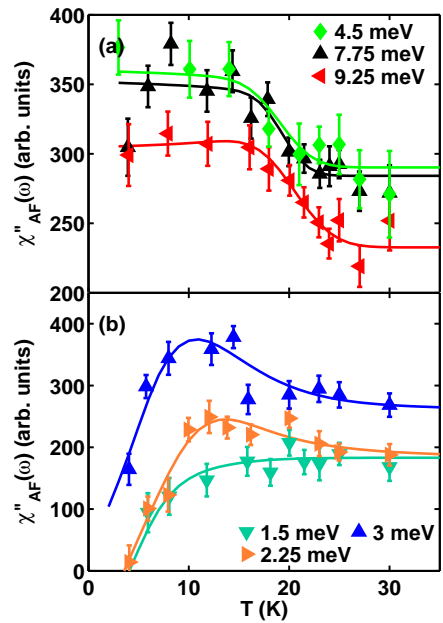}
\caption{Temperature dependence of the peak susceptibilities at (a) $\protect%
\omega = 9.25$, 7.75, 4.5 meV and (b) $\protect\omega = 3$, 2.25, 1.5 meV. The
result at 4.5 meV represents the combined statistics of data at 
$\protect \omega = 4$ and 5 meV obtained for a second crystal with the same $T_c$ and nearly identical composition. }
\label{fig4}
\end{figure}

Raman scattering results for both NCCO and Pr$_{2-x}$Ce$_{x}$CuO$_{4+\delta }
$ \cite{QazilbashPRB05} demonstrate a nearly linear decrease above $x\approx
0.15$ of the energy scales in $B_{1g}$, $B_{2g}$ 
and $A_{1g}$ symmetry. 
Unlike for the hole-doped compounds, the energy scales $B_{1g}$ and $B_{2g}$
symmetry are nearly identical, suggesting that in both cases one effectively
measures the maximum SC gap $2\Delta _{el}$ at the intersection of the
underlying Fermi surface with the AF Brillouin zone, in between nodal and
anti-nodal directions \cite{QazilbashPRB05,MatsuiPRL05}. 
The chemical composition of our main NCCO sample ranges from $x=0.15$ to
0.164, which corresponds to a range of about 3 meV of these two energy
scales, as indicated by the two vertical bands in Fig.~2. These ranges are
in good agreement with the positions of the extrema in the low-temperature
magnetic susceptibility. In order to quantify this observation of a
correspondence of magnetic and electronic energy scales, we describe the 4 K
data for $\chi _{AF}^{\prime \prime }(\omega )$ in Fig.~2(b) by the sum of
two terms: a step function centered at $\omega _{1}=\omega _{B_{1g}}$,
describing the electronic gap, and a Gaussian centered at $\omega
_{2}=\omega _{A_{1g}}$, describing the peculiar excitation below the gap
energy. The widths of the two features are fixed, chosen to correspond to a
(Gaussian) broadening due to the Ce inhomogeneity, and the fit therefore
contains only two adjustable parameters: the amplitudes of the Gaussian [$%
A_{G}=363(18)$ (a.u.)] and of the step function [$A_{S}=351(8)$ (a.u.)].
As can be seen from Fig.~2(b), $\chi _{AF}^{\prime \prime
}(\omega )$ in the SC state is described in an excellent fashion based on
the knowledge of electronic energy scales from Raman scattering and the sample
inhomogeneity from chemical analysis. A similarly good fit is obtained for
the difference data in Fig. 2(c). We note that a separate five-parameter fit
of the $T=4$ K data yields $\omega _{1}=6.4(3)$ meV, $\omega _{2}=4.5(2)$
meV, and a Gaussian broadening of $\Gamma =2.7(3)$ meV (FWHM).

The magnetic response upon entering the SC state consists of two
components: a broad rearrangement of spectral weight from low energies to
energies above $\omega _{1}$, and an additional new component centered at $%
\omega _{2}$. We offer two possible explanations for this second energy
scale: it could either be related to the observation of a non-monotonic $d$%
-wave gap \cite{QazilbashPRB05,MatsuiPRL05}, or it may be the magnetic
resonance. The non-monotonic $d$-wave gap is characterized by a maximum away
from the antinodal direction. Photoemission work on Pr$_{1-x}$LaCe$_{x}$CuO$%
_{4+\delta }$ near optimal doping suggests that the antinodal gap value is
approximately 80\% of the maximum gap, whereas our data give $\omega
_{2}/\omega _{1}=70(2)\%$. However, it seems unlikely that a significant
contribution to the low-energy magnetic response stems from the antinodal
regions, since they are not spanned by the wavevector 
$\mathbf{Q}_{AF}$ \cite{ArmitagePRL02}.

On the other hand, the low-energy feature may be the magnetic resonance. 
Figure 3(a) demonstrates that that a continuous enhancement of 
$\chi _{AF}^{\prime \prime }(\omega )$ upon cooling into the SC state
already exists around $\omega = 4.5$ meV. While our data are overall consistent with  
Refs.~\cite{WilsonNature06, ZhaoPRL07}, these earlier  results were interpreted as  
indicative of a resonance mode at $\omega_r = 9.5$-11 meV.
This conclusion was supported by the observation that the ratio 
$\omega_r/k_BT_c \approx 5$ is in good agreement with results for hole-doped cuprates.
However, recent work for Hg1201 revealed that the ratio $\omega_r/k_BT_c$ is not universal \cite{Guichuan08}. 
Furthermore, the interpretation of the susceptibility enhancement around 10 meV as the resonance is inconsistent with the fact that the resonance in the hole-doped compounds lies below $2\Delta _{el}$ \cite{EschrigAdvPhys06}. 

Recent STM work on Pr$_{0.88}$LaCe$_{0.12}$CuO$_{4+\delta }$ ($T_{\mathrm{c}%
}=24\,\mathrm{K}$) \cite{NiestemskiNature07}
indicated that $2\Delta _{el}\approx 14$ meV,
significantly larger than in NCCO \cite{Kashiwaya98,QazilbashPRB05}. Therefore, while our results are
inconsistent with the existence of a resonance near 10 meV in NCCO \cite%
{ZhaoPRL07}, they are not necessarily inconsistent with the original
observations of Ref.~\cite{WilsonNature06}. It will be important to confirm
the observation of a relatively large electronic gap in Pr$_{0.88}$LaCe$%
_{0.12}$CuO$_{4+\delta }$ with other experimental methods. Furthermore, 
STM revealed a bosonic mode at about 10.5 meV, consistent with the enhancement of magnetic susceptibility around 10 meV in Pr$_{0.88}$LaCe$_{0.12}$CuO$_{4+\delta }$  
\cite{NiestemskiNature07}.  
If the bosonic mode is indeed magnetic in origin, STM 
should locate it at $\omega _{2}$ in optimally-doped NCCO.

The interpretation of the $\omega_2$ feature as the resonance implies that the resonance energy agrees well with the Raman $A_{1g}$ response, consistent with the suggestion for the hole-doped compounds that the latter may be associated with the magnetic resonance \cite{Gallais2002, Letacon2005, Letacon2006JPCS}. However, this connection is not conclusive and the origin of the $A_{1g}$ peak is still unclear.

Figure 3(b) reveals a non-trivial non-monotonic temperature dependence at
lower energies. The enhancement of $\chi _{AF}^{\prime \prime }(\omega )$
at $\omega = 4.5$ meV
and the non-monotonic temperature dependece at $\omega = 2.25$ and 3 meV
appear to be the joint effect of the opening of the gap and the formation of
a new excitation below $2\Delta_{el}$ in the SC state. 
We note that the interpretation of the low-energy response in terms of a gap and an in-gap excitation requires a reinterpretation of the results of Refs.~\cite{YamadaPRL03,MotoyamaPRL06} in which the low-energy edge of the magnetic susceptibility in the SC state was directly associated with a gap.

Figure 2(e) shows the local susceptibility, $\chi ^{\prime \prime }(\omega
)=\int dQ^{3}\chi ^{\prime \prime }(Q,\omega )/\int dQ^{3}$. In the normal
state, $\chi ^{\prime \prime }(\omega )$ increases monotonically from zero
at $\omega =0$ meV to about $2.5\,\mu _B^2\,$eV$^{-1}$f.u.$^{-1} $ 
at 12 meV. In the SC state, $\chi ^{\prime \prime }(\omega )$
exhibits a local maximum of $\approx 2\mu _B^2\,$eV$^{-1}$f.u.$%
^{-1}$ at $\,\omega _{2}$ and reaches $\approx 4\,\mu _B^2\,$eV$%
^{-1}$f.u.$^{-1}$ at 12 meV. 
The estimated mean-square fluctuating moment $\langle m^{2}\rangle =\int
d\omega \,\chi ^{\prime \prime }(\omega )$ integrated up to 12 meV is 0.019
and 0.021 $\mu _B^2\,$f.u.$^{-1}$ at 30 and 4 K, respectively.
Although the difference is only 10\%, the data indicate that integration
somewhat beyond 12 meV would lead to a larger difference. Unfortunately, the
regime just above 12 meV is inaccessible due to high background scattering
from Nd$^{3+}$ crystal field excitations. This suggets that the
susceptibility enhancement above $\omega _{1}$ and the $\omega _{2}$ feature
cannot both be compensated for by the low-energy spectral weight loss in
the superconducting state, and that some high-energy (above 12 meV) spectral
weight must shift to lower energy to satisfy the total moment sum rule. If
the $\omega _{2}\approx 4$ meV feature were not present, the
superconductivity-induced local susceptibility enhancement above $\omega
_{1} $ ($\approx 0.004\,\mu _B^2\,$f.u.$^{-1}$ up to 12 meV)
would be fully covered by the spectral weight depletion in the gap ($\approx
0.007\,\mu _B^2\,$f.u.$^{-1}$). We estimate the momentum- and
energy-integrated weight of the low-energy feature to be 
$0.007(2)\mu _B^2\,$f.u.$^{-1}\,$ This value is somewhat smaller (by a factor
of 3-10) than the weight of the resonance in the hole-doped compounds 
\cite{EschrigAdvPhys06}.

While the magnitude of $\chi ^{\prime \prime
}(\omega )$ is consistent with work on Pr$_{0.88}$LaCe$_{0.12}$CuO$%
_{4+\delta }$ \cite{WilsonNature06,KrugerPRB07}, the SC and normal state
responses [both $\chi ^{\prime \prime }(\omega )$ and $\chi _{AF}^{\prime
\prime }(\omega )$] of the latter system do not approach zero at low
energies. 
The magnetic gap is an expected feature of the excitation spectrum, 
and it is also observed in hole-doped YBa$_{2}$Cu$_{3}$O$_{6+\delta }$ 
\cite{YBCOGap} and La$_{2-x} $Sr$_{x}$CuO$_{4}$ \cite{YamadaPRL95} near optimal
doping. 
We speculate that, as in early measurements on hole-doped 
La$_{2-x} $Sr$_{x}$CuO$_{4}$ 
\cite{EarlyLSCO}, disorder effects might mask the underlying low-energy
response of Pr$_{1-x}$LaCe$_{x}$CuO$_{4+\delta }$.

In summary, careful analysis of new neutron scattering data for
electron-doped Nd$_{2-x}$Ce$_x$CuO$_{4+\delta}$ reveals quantitative
agreement for the low-energy magnetic response at the antiferromagnetic wave
vector with electronic energy scales from Raman scattering. The larger of the two
magnetic energies corresponds to the electronic gap maximum, while the
smaller scale might possibly be the magnetic resonance. 

This work was supported by the DOE under Contract No. DE-AC02-76SF00515 and
by the NSF under Grant. No. DMR-0705086.

\bibliography{NCCO_PRL_Nov28}

\end{document}